\documentclass[11pt]{article}
\usepackage[square]{natbib}
\usepackage{authblk}
\usepackage{graphicx}
\usepackage{wrapfig}
\usepackage[a4paper,bindingoffset=0.2in,left=0.52in,right=0.85in,top=0.79in,bottom=1in,footskip=.25in]{geometry}
\usepackage{fancyhdr}
\usepackage[breaklinks,colorlinks = true,
            linkcolor = blue,
            urlcolor  = blue,
            citecolor = blue,
            anchorcolor = blue,
            breaklinks]{hyperref}
\usepackage{doi}
\usepackage{color}

\title{{\Large \bf Ionosphere-thermosphere global time response to geomagnetic storms}}
\date{}
\author[1,2]{{\large Denny M. Oliveira}}
\author[1]{{\large Eftyhia Zesta}}
\author[1]{{\large Peter W. Schuck}}
\author[1,3]{{\large Hyunju K. Connor}}
\author[4]{{\large Eric K. Sutton}}

\affil[1]{{\normalsize NASA Goddard Space Flight Center, Greenbelt, MD USA}}
\affil[2]{{\normalsize Goddard Planetary Heliophysics Institute, University of Maryland, Baltimore County, Baltimore, MD USA}}
\affil[3]{{\normalsize University of Maryland College Park, College Park, MD USA}}
\affil[4]{{\normalsize Air Force Research Laboratory, Kirkland AFB, Albuquerque, NM USA}}

\begin{document}

\maketitle

\begin{abstract}

In this study, we investigate thermospheric neutral mass density heating associated with 168 CME-driven geomagnetic storms in the period of May 2001 to September 2011. We use neutral density measured by two low-Earth orbit satellites: CHAMP and GRACE. For each storm, we superpose geomagnetic and density data for the time when the IMF B$_\mathrm{z}$ component turns sharply southward chosen as the zero epoch time. This indicates the storm main phase onset. We find that the average SYM-H index reaches the minimum of $-$42 nT near 12 hours after storm main phase onset. The Joule heating is enhanced by approximately 200\% in comparison to quiet values. In respect to thermosphere density, on average, high latitude regions (auroral zones and polar caps) of both hemispheres are highly heated in the first 1.5 hour of the storm. The equatorial response is presumably associated with direct equator-ward propagation of TADs (traveling atmospheric disturbances). A slight north-south asymmetry in thermosphere heating is found and is most likely due to a positive B$_\mathrm{y}$ component in the first hours of the storm main phase.
\end{abstract}

\thispagestyle{fancy}

\section{Introduction}

The ionosphere and the thermosphere are the most important regions of the Earth's upper atmosphere. The main populations of these layers are ions and neutral particles and molecules, respectively. Although these layers have distinct populations, they superpose each other in regions of low altitudes between 80-600 km altitude, where LEO (low-Earth orbit) satellites fly around the Earth. The ionosphere and thermosphere are mechanically connected to each other and the charged particles energize the neutral particles by collisions. During times of strong geomagnetic activity, the effects of this coupling become much more amplified \citep{Prolss2011}. \par

The main sources of energy for the upper atmosphere are the solar EUV (extreme ultra-violet) radiation, the Joule heating deposited there through very conductive magnetic field lines, and particle precipitation energy \citep{Prolss2011}. On average days, the EUV radiation energy input dominates over the electromagnetic energy inputs. However, during storm times, the electromagnetic energy input may correspond to almost 70\% of the total energy budget during intense events \citep{Knipp2004}. \par

The magnetosphere is strongly driven by solar wind forcing during storm times, when magnetospheric energy is transferred into the atmosphere. A myriad of events arise from this energy deposition. For example, auroral activity is intensified, ionospheric convection and conductivity increase which in turn facilitates the flow of strong electric currents connecting the magnetosphere and the ionosphere \citep{Zesta2000,Oliveira2014b}. This process, named Joule heating (JH), leads to the heating and upwelling of the neutral thermosphere \citep{Prolss2011}. As a result, LEO satellites may be subjected to larger atmospheric orbital drag forces which can reduce their life times and increase orbit prediction errors \citep[see, e.g., review by][]{Zesta2016b}. \par

In this study, we investigate the thermosphere global time response to geomagnetic storms. It is well known since pioneer investigations that the high latitude regions of the Earth's atmosphere are the first regions to undergo heating through storm forcing \citep{Prolss2011}. For example, \cite{Taeusch1971} showed that the thermosphere was heated in a time lag of 1.5 hour after storm onset. \cite{Burns1992} reported results of the magnetic equatorial thermosphere response corresponding to 3-6 hours after storm onset. Results of numerical simulations compared to observations as discussed by \cite{Connor2016} and \cite{Shi2017} show that the high latitude thermosphere was highly perturbed by the compression associated with an event with large dynamic pressure enhancement. \cite{Bruinsma2006} reported thermosphere heating in high latitude regions near 2 hours after an SSC (storm sudden commencement) event. These authors showed that low latitudinal response lagged the SSC onset by approximately 4 hours. \cite{Sutton2009a} found similar results as well. \par

As discussed above, many efforts have been made in the attempt of understanding the thermosphere response in a global context. However, a superposed epoch analysis on this subject was still lacking in the literature. The primary purpose of this study is to fill this gap. Therefore, we use 168 CME-caused geomagnetic storms in the time period of May 2001 to September 2011 to investigate spatial and temporal heating of the thermosphere neutral density in response to CME forcing. We use thermosphere neutral density as measured by accelerometers onboard the CHAMP (CHallenge Mini-satellite Payload) and GRACE (Gravity Recovery Climate Experiment) spacecraft. More details on this study can be found in \cite{Oliveira2017c}.

\section{Data and methodology}

\subsection{Data and instrumentation}

We use a data set of 168 geomagnetic storms caused by CMEs from May 2001 to September 2011. The times of arrival at Earth for these CMEs can be found in the CME catalogue located at \url{http://www.srl.caltech.edu/ACE/ASC/DATA/level3/icmetable2.htm#(a)}. As discussed by \cite{Oliveira2017c}, the number of storms agrees well with solar activity, result consistent with the occurrence of interplanetary shocks and solar activity \citep{Oliveira2015a}. More details on this data set have been discussed by \cite{Oliveira2017c}. \par

The 1-minute time resolution SYM-H index was downloaded from the World Data Center of Kyoto, Japan website \url{http://wdc.kugi.kyoto-u.ac.jp/aeasy/index.html}. The IMF data, with cadence of 1 minute, were downloaded from the NASA/Goddard Space Flight Center's OMNIWeb Plus website (\url{https://omniweb.gsfc.nasa.gov}). The IMF data have been shifted to the bow shock nose in order to be superposed along with other geomagnetic indices and density. The JH associated with the CME-caused storms were calculated by using the empirical model described by \cite{Knipp2004}. That model uses formulas corresponding to combinations of polar cap indices and the Dst index obtained from observations covering a time period of over two solar cycles. These formulas take into account season effects. The JH resolution is 1 hour. \par

The thermosphere neutral mass density data were obtained from accelerometers onboard two LEO satellites, CHAMP and GRACE. CHAMP was launched in September 2000 in an orbit with inclination of 87.25$^\circ$ at the initial altitude of 456 km \citep{Reigber2002}. The period of CHAMP was about 90 minutes.  CHAMP re-entered the Earth's atmosphere in September 2010. CHAMP covered all hours of local time in approximately 131 days. The time resolution of the CHAMP density data was 0.1 Hz. More details about the CHAMP data can be found in the literature \citep{Bruinsma2006}. \par

The GRACE constellation corresponds to two identical satellites, GRACE-A and GRACE-B, that were launched in March 2002 at the initial altitude of 500 km, in a quasi-circular orbit with inclination of 89.5$^\circ$ \citep{Tapley2004b}. GRACE-A is followed by GRACE-B in a spatial separation of 220 km. By the time of this writing, April 2017, both GRACE spacecraft are still in orbit around the Earth. The GRACE orbit period is 95 minutes. All hours of local time are covered by GRACE in a time interval of 160 days. The GRACE data have a finer resolution of 1 Hz. More details on the GRACE data can be found elsewhere \citep[see, e.g.,][]{Bruinsma2006}. 

\subsection{Methodology}

Since the CHAMP and GRACE spacecraft correspond to LEO spacecraft, their altitudes undergo slight variations during consecutive orbits. Such small variations may cause differences in density measurements obtained in the thermosphere dayside and nightside, where the dayside density is larger due to solar EUV heating. In order to minimize this altitude effect, all the density data were normalized to a common altitude of 410 km according to the well-known Jacchia-Bowman model, version 2008 (JB2008). The model is described in details by \cite{Bowman2008}. The relationship used to calculate the normalized density is given by:
\begin{equation}
\rho_\mathrm{410}=\frac{\rho_{JB410}}{\rho_{JB}}\times\rho\,,
\end{equation}
where $\rho$ represents the raw density measured by either CHAMP or GRACE at each spacecraft altitude, $\rho_\mathrm{JB}$ is the density obtained by JB2008 at the satellite altitude, and $\rho_\mathrm{JB410}$ is the density calculated by JB2008 at the specific altitude of 410 km, and $\rho_{410}$ is the normalized neutral mass density at h = 410 km. \par

In order to obtain a sense of neutral density variation during storm times, we calculate the background density during quiet times using the JB2008 model. As described by \cite{Bowman2008}, the JB2008 model depends on a single parameter, the local exospheric temperature T$_\infty$, to estimate the neutral density in the thermosphere. This is computed by $T_\infty=T_\ell(\theta,\delta_\odot,H)+\Delta T_{LST}(H,\theta,z)+\Delta T_{Dst}$ where $T_\ell$ is an empirical formula for the local exospheric temperature based on latitude $\theta$, solar declination $\delta_\odot$ and solar local time $H$, $\Delta T_{LST}$ is a height-dependent local solar time correction, and $\Delta T_{Dst}$ is a global correction due to geomagnetic activity computed from the Dst index. We then obtain the quiet density by the model input with no Dst correction, or $\Delta T_{Dst}$ = 0, for the projected altitude of 410 km, denoted $\rho_{\mathrm{Q, 410}}$. See more details about this approach in \cite{Oliveira2017c}. \par

Therefore, in order to capture spatial and temporal variations of the thermosphere neutral mass density during storm times, we compute and superpose the following dimensionless quantity, the logarithm of the density gain factor, the ratio of the normalized density to the quiet density at 410 km according to the JB2008 model:

\begin{equation}
\log_{10}\left[ \frac{\rho_{410}}{\rho_{\mathrm{Q, 410}}} \right]
\end{equation}

\section{A case example: the 24 August 2005 geomagnetic storm}

\begin{figure*}[t]
\centering
\vspace*{-0.1cm}
\includegraphics[width=0.72\textwidth]{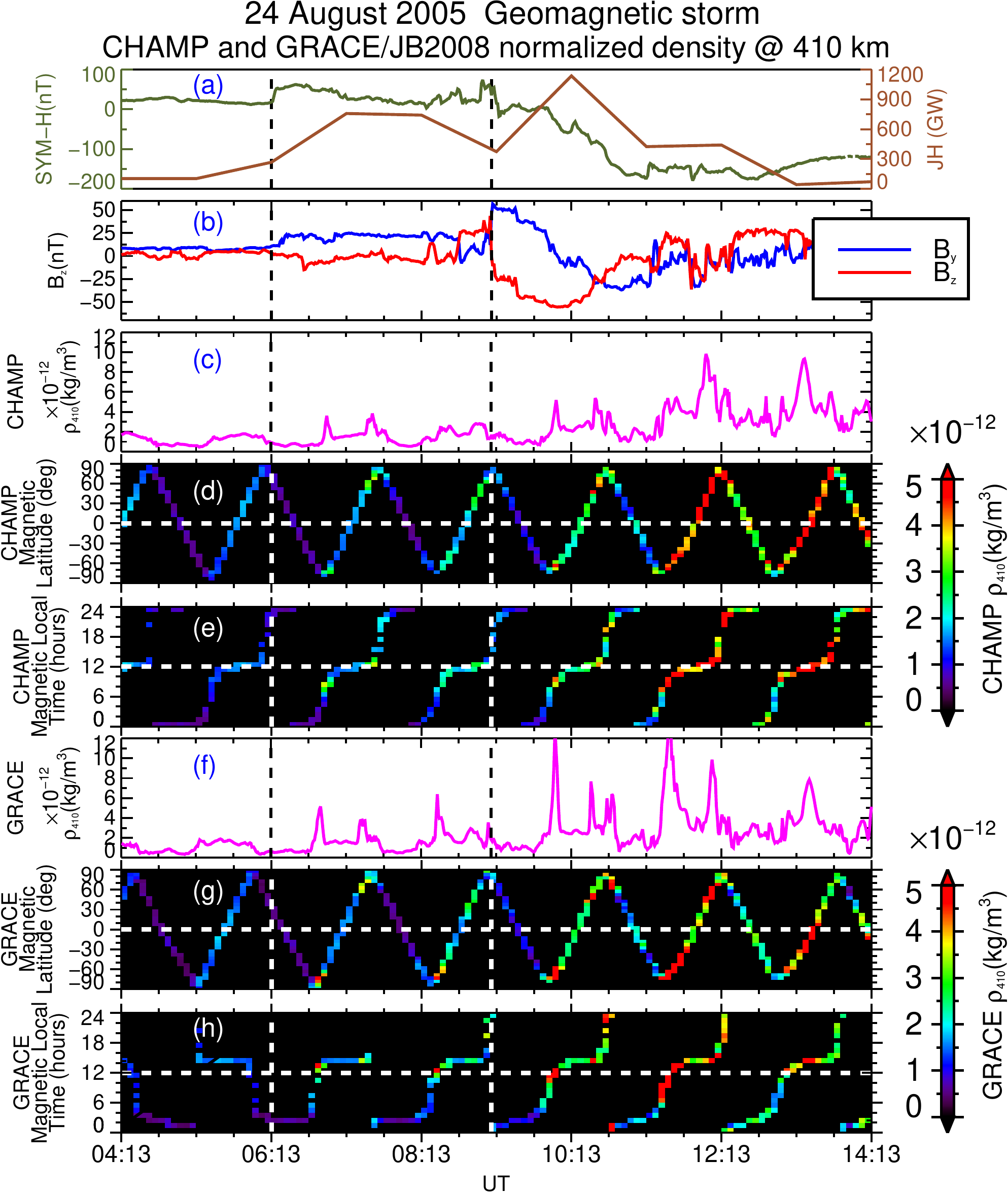}
\vspace*{-0.4cm}
\caption{{\footnotesize Summary plot of the 24 August 2005 geomagnetic storm. From top
  to bottom: (a), SYM-H (green) and JH (brown); (b), IMF B$_\mathrm{y}$ (blue) and B$_\mathrm{z}$ (red); (c), CHAMP density as a
  function of: (c), UT; (d), MLAT and UT; and (e), MLT and UT. Panels (f),
  (g) and (h) show the same for GRACE. The vertical dashed lines
  indicate: first, CME impact, and second, storm main phase onset. The CHAMP and GRACE densities were normalized to the common altitude of 410 km according to the JB2008 model.}}
\label{event}
\end{figure*}

In order to illustrate the methodology used in this study, we choose the 24 August 2005 geomagnetic storm event as a case study example. IMF, geomagnetic indices and neutral density data for that event are summarized by Figure \ref{event}. Figure \ref{event}(a) shows the SYM-H index (green), in nT, and the Joule heating JH (brown), in GW, as calculated by empirical formulas derived by \cite{Knipp2004} with season effects accounted for; (b) shows B$_\mathrm{y}$ and B$_\mathrm{z}$ in blue and red colors, respectively. The normalized thermosphere densities $\rho_{410}$ for the CHAMP and GRACE satellites are indicated by panels (c) and (f), respectively. Panels (d) and (g) show the same densities averaged, color coded and allocated in 5$^\circ$-bins in magnetic latitude (MLAT) and 5-minute bins in time for CHAMP and GRACE, respectively. Finally, panels (e) and (h) show the same for CHAMP and GRACE in magnetic local times (MLT) bins and the same bin size for time. \par

In all panels, the first dashed vertical lines indicate the time of CME impact (according to the CME list). An SSC event is clearly shown by an abrupt jump of the SYM-H index at 0613 UT. The second dashed vertical line indicates the moment of the southward turning of the IMF B$_\mathrm{z}$ component. Approximately 3 hours later, at 0909 UT, SYM-H starts to undergo a sharp decrease indicating the storm main phase onset. In the next two hours, SYM-H reaches minimum values around $-$180 nT and stays at that level up to $\sim$1330 UT, when the magnetosphere starts to recover. At 0909 UT, after a sharp inversion, B$_\mathrm{z}$ reaches a minimum value of $-$50 nT during approximately 1 hour when it starts to increase to positive values, eventually oscillating between null values after 1130 UT. Similarly, B$_\mathrm{y}$ jumps at 0909 UT from slightly positive values to a maximum value over $+$50 nT, becoming moderately negative ($-$25 nT) in the next hour or so and then increases to null values, oscillating there in the subsequent hours. Finally, JH shows two distinct enhancements: first, between 0613 and 0909 UT when it increases from 30 GW to 600 GW, and then after 0909 UT when it increases from 400 GW values to values as high as 1200 GW. The first JH enhancement is probably due to a combination of shock/compression \citep{Connor2016,Shi2017} and positive B$_\mathrm{y}$ \citep{Knipp2011,Li2011} effects, while the second enhancement is associated with the storm main phase effects. \par

The plots representing density data show that both CHAMP and GRACE executed orbits around the Earth in very similar latitudes and slightly different local times for that event, with the former in lower altitudes than the latter. Figures \ref{event}(c) and (g) clearly show two density peaks for CHAMP and GRACE at approximately 0700 and 0745 UT, whereas GRACE showed a larger peak at 0845 UT not observed by CHAMP. These density enhancements are presumably associated with the CME impact. After the storm main phase onset, several peaks are observed for both satellites every 45 minutes due to the deposition of magnetospheric energy into the upper atmosphere. \par

The remaining panels show very similar results for both CHAMP and GRACE, and therefore will be described in a common way as follows. Before shock impact, density varies slightly in the dayside equatorial region due to EUV heating. After 0613 UT, density enhancements due to compressions are observed in the first southern polar pass at 0650 UT. Such perturbations are more intensely observed in the subsequent equatorial passes and the first northern polar pass. Compression effects are still seen in the following orbits, but perturbations fade away in the first descending orbit after 0909 UT. Then, at approximately 1000 UT, time of the first southern pass during storm main phase, densities are observed to be more intensified in comparison to those caused by the compression effects. As a result, approximately 3 hours after storm main phase onset, thermosphere density is heated in the equatorial regions. Since these perturbations propagate from high to low latitude regions, they are most likely associated with TADs (traveling atmospheric disturbances) \citep{Bruinsma2007,Sutton2009a}. Then, near 1500 UT (not shown), the magnetosphere starts to recover from the storm and the thermosphere starts to cool off reaching similar density values as those seen in the pre-storm state. \par

\begin{figure*}[t]
\centering
\includegraphics[width=0.72\textwidth]{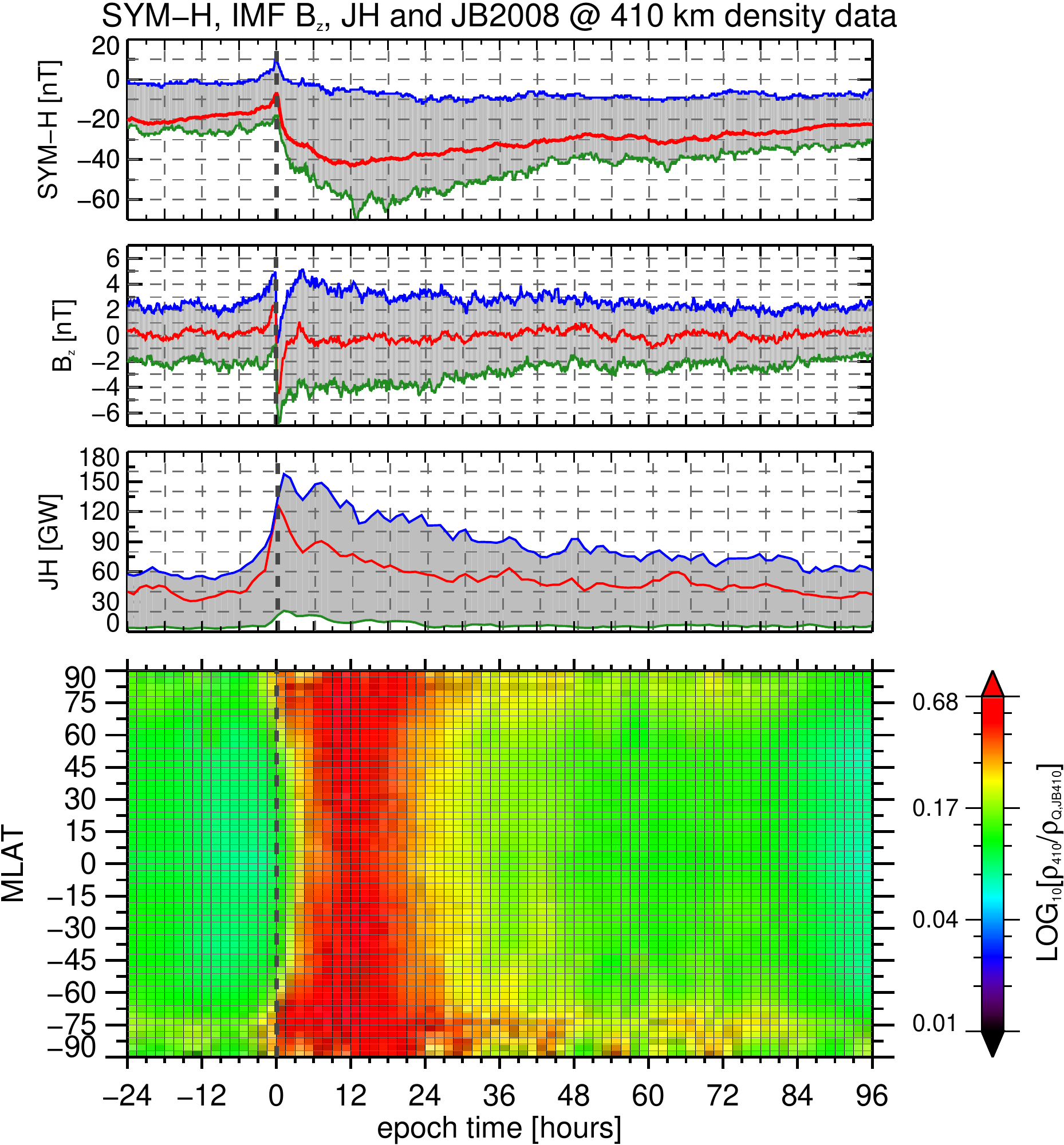}
\vspace*{-0.5cm}
\caption{{\footnotesize First panel: the SYM-H index; second, the IMF B$_\mathrm{z}$ component, and third panel, Joule heating. The thick red lines indicate average values, the blue lines indicate the 75$^\mathrm{th}$ percentile, and the green lines indicate the 25$^\mathrm{th}$. In the last panel, $\log[\rho_{410}/\rho_{Q, JB410}]$ is color coded an plotted as a function of MLAT and epoch time. The size of each grid is 3$^\circ\times$90 minutes.}}
\label{special}
\end{figure*}

CHAMP and GRACE executed orbits mostly in the local times of 1200 hour in the dayside and 2300-2400 hours in the nightside. In the moments preceding the shock impact, thermosphere density showed slight heating due to solar heating around 1200 UT. After the compression, the thermosphere is heated mostly in the dayside, showing a slight nightside heating in the last orbit before storm main phase onset. In the second equatorial pass during storm main phase onset, the thermosphere is highly heated in the dayside and it is seen to spread out to all local times in the subsequent hours. However, the heating is stronger in the dayside. \par

This case example, explained in details, indicates that the thermosphere can show broad response not only in latitude, but also in local time to storm forcing. Therefore, the thermosphere heating during geomagnetic storms is a global phenomenon. In the next section we will show results of the superposed epoch analysis of 168 CME-caused geomagnetic storm for the storm main phase onset chosen as the zero epoch time for all events.

\section{Results}

As seen with details in the description of Figure \ref{event}, energy input from the CME interaction with the magnetosphere produces two distinct effects on the thermosphere. First, the heating of the thermosphere due to the CME impact, and, second, the heating caused by the arrival of the CME magnetic structure at the Earth. The CME impact is usually associated with an SSC event whose signature is a sharp increase in the SYM-H index. That is the storm initial phase. When the magnetic structure interacts with the geomagnetic field, the southward B$_\mathrm{z}$ component reconnects with geomagnetic \begin{wrapfigure}{r}{0.52\textwidth}
\centering
\vspace*{-0.2cm}
\includegraphics[width=0.50\textwidth]{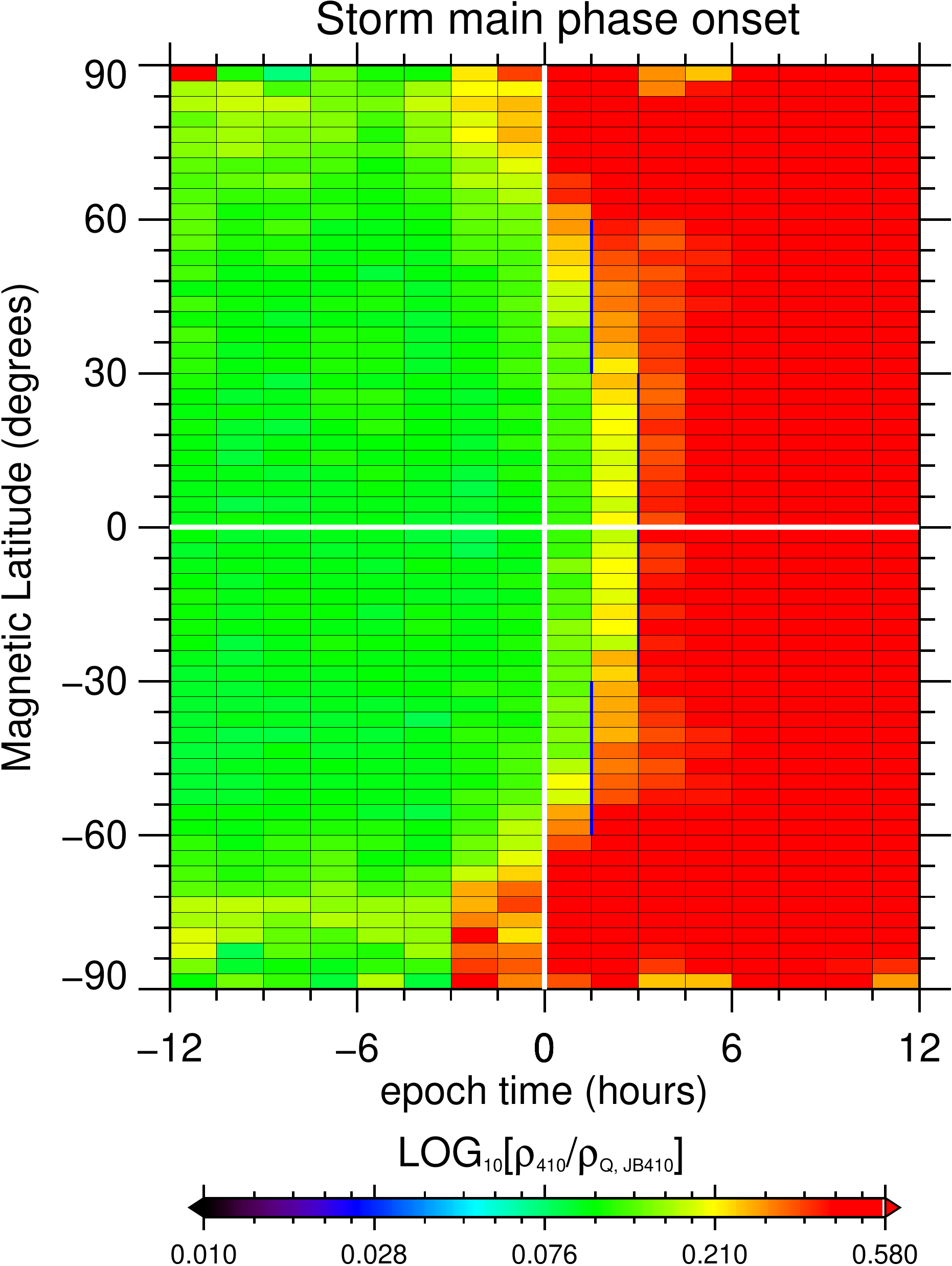}
\vspace*{-0.4cm}
\caption{{\footnotesize Thermosphere neutral density plotted in the same way as in Figure \ref{special}, but in a time interval of 12 hours before and after zero epoch time.}}
\vspace*{-0.3cm}
\label{special_blow-up}
\end{wrapfigure} field and large amounts of energy and momentum gain access to the magnetosphere and subsequently the upper atmosphere. This is usually associated with the storm main phase onset, when the SYM-H index is highly depressed due to storm activity.  As described by \cite{Oliveira2017c}, the time interval of interaction between these two CME regions are typically smaller than 1 hour, but some events can have time intervals as large as 6 hours. \cite{Oliveira2017c} detailed the differences between the effects of these two CME regions on the thermosphere when the CME shock impact and the IMF B$_\mathrm{z}$ southward turning were taken as zero epoch times. In this paper, however, we will focus on the effects caused by the geomagnetic structure on thermosphere heating during storm times. \par

The zero epoch time for each event was chosen, by inspecting the OMNI data visually, as the time in which the IMF B$_\mathrm{z}$ turns abruptly southward. Then, for each event, SYM-H, IMF B$_\mathrm{z}$ and JH data were lined up 24 hours before and 96 hours after the storm main phase onset for each event. \par

Figure \ref{special} shows the results obtained for this superposition. The first three panels of that figure show the SYM-H index, in nT; IMF B$_\mathrm{z}$, in nT; and JH, in GW. All these quantities were superposed and averaged every 5 minutes. The thick red lines indicate averaged values, green lines indicate lower quartiles, and blue lines indicate upper quartiles for each quantity. The vertical black dashed lines indicate the zero epoch time. \par

The bottom panel of Figure \ref{special} shows the histogram of the logarithm of the gain factor, $\log_{10}[\rho_{410}/\rho_{\mathrm{Q, 410}}]$, as a function of MLAT and epoch time. The size of each grid is (MLAT, epoch time) = 3$^\circ$$\times$1.5 hour. This is the latitudinal resolution of the data. Each grid represents one satellite orbit in time. The black dashed vertical line indicates the zero epoch time. \par

The top panel of Figure \ref{special} shows that, on average, at t = 0, the SYM-H index drops from values of$-$9 nT to the minimum value of $-$42 nT, which corresponds to the main phase of an average storm with duration of near 12 hours. The next panel shows that the averaged B$_\mathrm{z}$ component underwent an abrupt inversion from +2 nT to $-$5 nT at the storm main phase onset. B$_\mathrm{z}$ then returns to almost null values in about 6 hours. In the case of JH, as shown in the third panel, average values start to increase before zero epoch times due to the CME impacts and compressions preceding the storm main phase by at least a few minutes \citep{Connor2016,Shi2017}. This effect was seen in Figure \ref{event} for a particular event. Overall, JH increases from 45 GW to 125 GW, an increase of near 200\%, and then decreases with little variations to previous values of 30 GW. \par

As seen in Figure \ref{special}, the first regions to respond to storm forcing are the high latitude regions, or the auroral zones and the polar cap, with MLAT approximately between $\pm$63$^\circ$ an $\pm$90$^\circ$ in the Northern and Southern Hemispheres. Mid-latitude regions are heated within the first orbit, and the equatorial regions are perturbed in a time frame of 3 hours or two satellite orbits. This is most likely due to TADs propagating from high to low latitude regions \citep{Bruinsma2007,Sutton2009a}. High latitude regions are more intensely heated after the fourth orbit, or 6 hours in both hemispheres. The thermosphere cooling starts to take place in mid-latitude regions and slightly earlier in low-latitude regions at about t = 21 hours. This cooling might be related to the competition of storm heating and Nitric Oxide (NO) cooling in those regions, which is the well-known ``thermostat" effect \citep{Knipp2017a}. The cooling in the high latitude regions takes longer due to probably two reasons: the later heating of and a lesser thermostat efficiency in those regions. \par

The thermosphere heating in equatorial regions is emphasized by Figure \ref{special_blow-up}. That figure shows the density gain factor plotted in the same way and grids as in Figure \ref{special}, but now from 12 hours before and after storm main phase onset. Heating before t = 0 is evident in both hemispheres, probably resulting from CME impacts before storm onset. As discussed above, high latitude regions are highly heated in the first orbit after storm onset. Mid-latitude regions are heated after the first orbit, and equatorial regions in the beginning of the third orbit. As can be seen with more details in this figure, the Southern Hemisphere response is slightly higher in comparison to the Northern Hemisphere. This inter-hemispheric asymmetry is probably related to an asymmetry in the B$_\mathrm{y}$ component, as suggested by \cite{Yamazaki2015a}. This thermosphere feature will be investigated in a further work.

\section{Conclusion}

In this work, we investigated the global time response of the thermosphere neutral mass density to 168 CME-caused geomagnetic storms. These storms covered a time period of over 10 years, or almost one solar cycle. Neutral mass density for these events were represented by data obtained by state-of-the-art accelerometers carried by the CHAMP and GRACE spacecraft. In order to minimize altitude effects, all the neutral mass density data were projected to the common altitude of 410 km as computed by the JB2008 model. Background and solar activity effects were eliminated as much as the model allows by calculating the background density at 410 km and computing the variation of the density gain factor as $\log_{10}[\rho_{410}/\rho_{Q, JB410}]$. \par

We superposed SYM-H, IMF B$_\mathrm{z}$ and Joule Heating data with the IMF B$_\mathrm{z}$ component turning southward time for each storm taken as the zero epoch time. We found that, on average, the SYM-H index decreases at t = 0, B$_\mathrm{z}$ undergoes an abrupt change from null-positive to negative values, and JH goes from approximately quiet values of 30 GW to 120 GW, which accounts for a variation of near 300\%. \par

In respect to the neutral density, some thermosphere heating is found in the high latitude regions of both hemispheres due to CME impacts preceding the onset of the storm main phase. Right in the first orbit, at t = 1.5 hours, density starts to increase in the auroral zones and polar caps in both hemispheres due to large energy and momentum deposition there. In the next orbit, within t = 3.0 hours, perturbations reach the equatorial regions, mostly carried directly by TADs, as previously reported \citep{Bruinsma2007,Sutton2009a}. Mid latitude regions ($\pm$30$^\circ$-60$^\circ$ MLAT) starts to recover a few orbits ($\sim$6 hours) before the equatorial thermosphere. The high latitude regions, however, take longer than the other regions to recover probably due to two factors: later perturbations associated with the arrival of CME structures at the magnetosphere and a lesser ``thermostat'' effect in those regions since the storms were preceded by shocks and/or compressions \citep{Knipp2017a}. More details about this study can be found in \cite{Oliveira2017c}. \par

Finally, since most thermosphere models use lag times greater than the 3.0 hour  lag times for global response found in this study, we suggest some thermosphere models should be reviewed in order to improve their neutral mass density computation performance.

\setlength{\bibsep}{0.1pt}

{\small

}

\end{document}